\documentstyle[aps,prl,twocolumn,graphics]{revtex}

\newcommand{\ra}{\rangle}
\newcommand{\la}{\langle}
\newcommand{\om}{\omega}
\newcommand{\eps}{\epsilon}

\begin{document}
\draft
\wideabs{
\title{\bf Manipulation of Cold Atomic Collisions by Cavity QED Effects.}
\author{J.I. Kim, R.B.B. Santos, and P. Nussenzveig}
\address{Instituto de F\'\i sica, Universidade de S\~ao Paulo, {\protect \\}
Caixa Postal 66318, CEP 05315-970, S\~ao Paulo, SP, Brazil.}
\date{Received 10 May 2000}
\maketitle
\begin{abstract} 
We show how the dynamics of collisions between cold atoms can be 
manipulated by a modification of spontaneous emission times. This 
is achieved by placing the atomic sample in a resonant optical cavity. 
Spontaneous emission is enhanced by a combination of multiparticle 
entanglement together with a higher density of modes of the modified 
vacuum field, in a situation akin to superradiance. A specific situation 
is considered and we show that this effect can be experimentally observed 
as a large suppression in trap-loss rates.
\end{abstract}

}

Experiments with cold and ultracold atoms have led to many recent 
achievements, such as Bose-Einstein condensation~\cite{bec}, atom 
optics and interferometry~\cite{aoptics}, and precision 
measurements~\cite{precmeas}. Atomic collisions at these very low 
temperatures~\cite{weiner99} are of great importance in many of 
these applications. The density of atoms attainable in optical 
traps is usually limited by exoergic inelastic collisions, which 
lead to trap loss. The study of these processes presents several 
interesting features, since the dynamics is very distinct from 
collisions at higher temperatures. Because the atoms move so 
slowly, not only are they sensitive to long-range interaction 
potentials but, in the presence of light, they can also undergo 
changes of internal states during a collision (the interaction 
time can be larger than the typical spontaneous emission time). 
Therefore, these collisions can be manipulated with light, as 
demonstrated, for example, in experiments of optical 
shielding~\cite{shielding} of trapped atoms from collisional 
loss. 

One parameter, however, has been overlooked for the optical 
manipulation of cold collisions. The final outcome of a two-body 
encounter depends strongly on the spontaneous emission time. In 
this Letter we show, for the first time, how to modify 
spontaneous emission times in the context of cold collisions, and 
thereby manipulate the collisional dynamics. This is done by a 
combination of multiparticle entanglement~\cite{entangl} together 
with a modified vacuum field in Cavity QED~\cite{qed}, in a 
situation akin to superradiance~\cite{dicke,bonifacio,superrad}. We focus here 
on one specific collision process and analyze it in the 
presence of an optical cavity. Orders of magnitude of trap-loss probabilities
show that this modification in collisional dynamics is within 
reach of current experimental techniques. 

\begin{figure}
\centering \resizebox{7.5cm}{!}{\includegraphics*{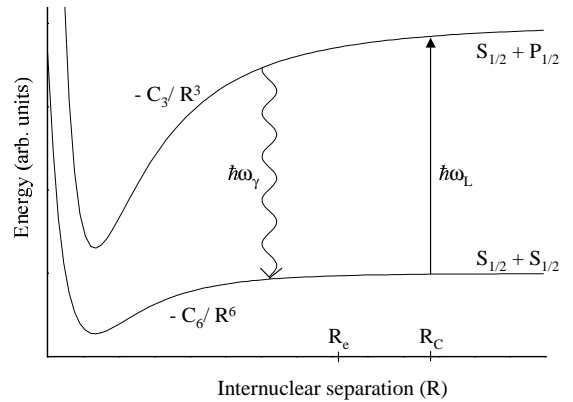}}
\caption{Excited state long range dipole-dipole potential 
$U=-C_3/R^3$ and the ground state van der Waals attractive 
potential $-C_6/R^6$, $C_6$ being a constant. Its distance to 
the asymptote of $U$ is the atomic separation $\hbar\om_A$ 
between $nS_{1/2}$ and $nP_{1/2}$.} \label{fig:1}
\end{figure}

Let us describe briefly one of the first identified collisional loss 
processes, so-called radiative escape from a trap~\cite{weiner99}. 
One atom of a colliding pair is excited by a laser of frequency $\om_L$ 
at a large internuclear separation 
$R$, and the atoms are accelerated towards each other by the strong 
long-range dipole-dipole attractive potential $U=-C_3/R^3$, where $C_3$ 
is a constant that depends on the atom under consideration. If the 
spontaneous emission time is long, the pair may gain enough kinetic energy 
to escape from the trap (by emitting a photon with energy 
$\hbar\om_{\gamma}$ smaller than that of the absorbed photon, $\hbar\om_L$). 
The interaction potentials are sketched in Fig.~\ref{fig:1} for the 
ground state $nS$ and the excited state $nP$ of an 
alkali atom. Two regions can be defined: $R<R_e$ and $R_e<R<R_C$, where 
$R_C$ (the so-called Condon point, chosen by tuning $\om_L$) is 
the internuclear separation at which a weakly bound molecule is excited. 
The separation $R_e$ is the smallest one for which spontaneous emission does 
not lead to trap loss. If one can {\it enhance} spontaneous emission of 
this excited molecule, decay will happen earlier, in the region between 
$R_C$ and $R_e$ in Fig.~\ref{fig:1}, preventing atoms from being lost. 

The modification of atomic radiative properties was one of the first effects to
be demonstrated in Cavity QED~\cite{qed}. Spontaneous emission
enhancement~\cite{decay,feld} and inhibition~\cite{kleppner85} were demonstrated 
in the 80's. Radiative level shifts, such as a cavity-induced Lamb shift, were  
also demonstrated in this context~\cite{feld,paulo94}. However, spontaneous
emission for single atoms in the optical domain was not significantly enhanced,
owing to the relatively small solid angle encompassed by a centimeter-sized
Fabry-Perot cavity. It is, nevertheless, possible to achieve a large enhancement
of spontaneous emission when we consider a sample of many identical weakly bound
molecules (so-called {\em quasimolecules}~\cite{weiner99}) coupled to the same 
cavity mode. The different excited quasimolecules are indistinguishable when
interacting with the cavity field. Quantum interference will thence be important
in the process of spontaneous emission into the cavity, in a way analogous to
superradiance~\cite{dicke,bonifacio,superrad}. Since, for most experiments in
cold collisions, the typical separation between different quasimolecules is
greater than an optical wavelength, this interference will only be constructive
if the quasimolecules are excited into a multiparticle entangled state by a
laser beam injected into the cavity mode~\cite{bonifacio}. Cavity QED effects on
cold atoms have been recently investigated for high-$Q$ cavities interacting
with single pairs of atoms in the context of cold collisions~\cite{deb99}, and
with large numbers of atoms in the context of forces exerted on the
atoms~\cite{gangl00}. 

\begin{figure}
\centering \resizebox{7cm}{!}{\includegraphics*{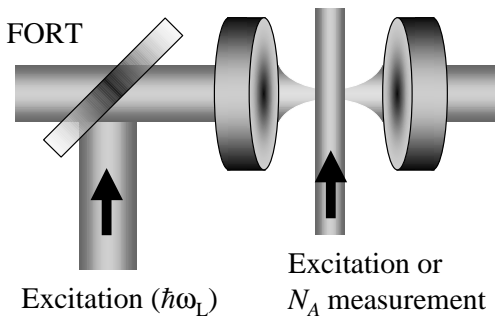}} 
\caption{Sketch 
of a possible experiment to probe Cavity QED-modified cold 
collisions. The $N_A$ atoms are held near the center of a Fabry-Perot 
cavity by a FORT laser. The excitation (probe) laser, of 
frequency $\om_L$, can be sent either perpendicular to the cavity 
axis or via the cavity mode.} \label{fig:2}
\end{figure}

We consider a sample of cold atoms trapped near the center
of a quasi-confocal optical resonator. In order to avoid optical pumping 
effects, atoms can be trapped in the ground state. This can be done,
for instance, if the atoms are in a far-off resonance trap (FORT)~\cite{fort},
as sketched in Fig.~\ref{fig:2}, or in a magnetic trap~\cite{magtrap}. The
experiment would require highly reflective mirror coatings with a sharp edge, in
order to transmit the FORT beam (detuned by a few nanometers from the atomic
transition). However, the whole setup would be far from trivial, since it is not  
straightforward to load a FORT (from a Magneto Optical Trap, MOT~\cite{mot}) 
inside an optical cavity in view of alignment difficulties. The dipole-dipole 
potential $U$ is turned on by an excitation laser
red detuned from the atomic transition (separation $\hbar\om_A$ between states 
$nS_{1/2}$ and $nP_{1/2}$). Our orders of magnitude will be
calculated for $^{85}$Rb atoms, so the atomic ground and excited states
in the following will be $5S_{1/2}$ and $5P_{1/2}$, respectively.

A colliding pair subject to a weak excitation field is likely 
to undergo just one-photon processes. The pairs are treated in the 
two-level approximation (which is justified later in the text)
with a ground state $|g\ra$ and an excited 
state $|e\ra$ connected to the $5S_{1/2}+5S_{1/2}$ and 
$5S_{1/2}+5P_{1/2}$ asymptotic states, respectively. 
Here, the pair in state $|g\ra$ is not bound but, in state 
$|e\ra$, there is a weak binding force. For simplicity, we will 
refer to these pairs as quasimolecules, independent of their 
state ($|e\ra$ or $|g\ra$). For a given $R$, the  energy 
separation between 
$|g\ra$ and $|e\ra$ is $\hbar\omega_R=\hbar\omega_A-C_3/R^3$ 
(Fig.~\ref{fig:1}).  Each quasimolecule interacts with the 
electromagnetic field with a dipole moment 
$(\sigma_i+\sigma_i^\dagger)\,{\bf d}_i$, where $\sigma_i$ and 
$\sigma_i^\dagger$ are Pauli operators acting in the subspace 
spanned by $|e\ra$ and $|g\ra$. The dipole ${\bf d}_i$, 
determined by the molecular axis~\cite{napolitano97}, is randomly 
oriented with respect to the cavity field polarization. Its 
magnitude is $|{\bf d}_i|=\sqrt{2}\,d_A$, where $d_A$ is the 
atomic dipole moment, with a resulting decay constant $\Gamma$ 
which is twice that of the atomic excited state 
$\Gamma_A$~\cite{weiner99}. 

When the quasimolecules are excited via the cavity mode, they end up  
in a multiparticle entangled state. We will only treat here the 
simplest entangled state, produced when a {\it single} excitation 
is injected into the cavity. In the ground state $|G;1_{{\bf 
k}_L}\ra\equiv |gg\cdots g;1_{{\bf k}_L}\ra$ all 
quasimolecules are in state $|g\ra$ and there is one laser photon in the cavity.
The quasimolecule--field 
interaction couples this state to all singly-excited states 
$|i;0\ra$ ($|1;0\ra\equiv |eg\cdots g;0\ra$, $|2;0\ra\equiv 
|ge\cdots g;0\ra$, and so on). 
Using the dipole and rotating-wave approximations~\cite{qed}, the 
matrix elements of the interaction hamiltonian 
$H_{\mbox{\scriptsize int}}$ are given by $V_i=\la 
i;0|H_{\mbox{\scriptsize int}}|G;1_{{\bf k}_L}\ra={\cal E} 
(\om_L) f^c({\mathbf r}_i)\mbox{\boldmath $\eps$}_L \cdot 
\mbox{\bf d}_i$. They depend on the positions ${\mathbf r}_i$ of 
the quasimolecules along the cavity mode, characterized by a 
profile $f^c({\mathbf r})$, polarization $\mbox{\boldmath 
$\eps$}_L$ and field per photon ${\cal 
E}(\om)=(2\pi\hbar\om/{\cal V})^{1/2}$, where ${\cal V}$ is the 
effective mode volume. The entangled state produced is finally 
given by 
\hbox{$|E;0\ra$}\hbox{$=\sum_iV_i|i;0\ra/\hbar\tilde\Omega$}, 
defining the collective Rabi frequency 
\hbox{$\tilde\Omega$}\hbox{$=(\sum_i|V_i|^2)^{1/2}/\hbar$}. Note 
that although a single excitation is present, it is shared by 
{\em all} pairs. 

We now calculate the spontaneous emission rate $\Gamma_c$ for this 
entangled state. Summing over the final states $|G;1_{\bf k}\ra$ and 
considering all possible wavevectors ${\bf k}$ for the emitted photon, 

\begin{equation}
\frac{\Gamma_c}{2\pi}=\int d\om_{\bf k}d\Omega_{\bf k}\ \rho\
\delta(\om_{\bf k}-\om_R)\, \frac{\sum_\lambda|{\bf X_k}\cdot\mbox{\boldmath
$\eps$}_\lambda|^2}{\hbar^2} \; ,
\label{eq:xk}
\end{equation}

\noindent 
where ${\bf X_k}\equiv\left(\sum_i{\cal E}(\om_{\bf 
k})f_{\bf k}({\bf r}_i)V_i^{\ast}{\bf 
d}_i/\hbar\tilde{\Omega}\right)$ and $f_{\bf k}({\bf r})$ is the 
mode function for a given ${\bf k}$. The product $|{\bf 
X_k}\cdot\mbox{\boldmath $\eps$}_\lambda|$ is the absolute value 
of the collective coupling between the quasimolecules and the 
field (matrix element of $H_{\mbox{\scriptsize int}}$). Taking 
into account that only emission into the solid angle 
$\Delta\Omega_c$ encompassed by the cavity mirrors is affected by 
the enhanced spectral density $\rho(\om)=\rho_0(\om)\Lambda(\om)$ 
(for the degenerate longitudinal 
modes), we separate its contribution from emission into the rest 
of free space. Here, $\rho_0(\om)$ is the free space density and 
$\Lambda(\om)$ is the cavity line shape function~\cite{density}. 
In view of the large number of degenerate transverse modes in the 
cavity, the integral over $d\Omega_{\bf k}$ can be replaced by a 
summation over transverse TEM$_{nm}$ modes with profiles 
$f_{nm}({\bf r})$~\cite{density}. The solid-angles 
$\Delta\Omega_{nm}$ encompassed by them are determined by their 
transverse dimensions at the mirrors. Diffraction losses are also 
accounted for, by substituting effective $\Lambda_{nm}$ values for 
each TEM$_{nm}$ mode. This correction is significant only for 
high-order modes, for which $\Delta\Omega_{nm}\sim 
\Delta\Omega_c$. An effective solid angle can be defined by 

\begin{equation}
\Delta\Omega_{eff}\equiv
\sum_{nm}\Delta\Omega_{nm}\frac{\Lambda_{nm}}{\Lambda_{00}}
\frac{|\frac{1}{N}\sum_i|\mbox{\boldmath $\eps$}_L\cdot{\bf
d}_i|^2\,f^\ast_{nm}({\bf r}_i)f^c({\bf
r}_i)|^2}{\frac{1}{N}\sum_i|\mbox{\boldmath $\eps$}_L\cdot{\bf
d}_i|^2 |f^c({\bf r}_i)|^2} \;, \label{eq:omegaeff}
\end{equation}

\noindent 
where $N$ is the number of entangled quasimolecules. 
For typical optical cavities, this solid angle will be relatively 
small. Emission into the rest of free space will then be little 
affected by the presence of the cavity and it can be calculated to be
approximately $\Gamma$~\cite{details}. For resonant excitation
$\om_R\sim\om_L$ ($|\om_R-\om_L|\lesssim\gamma_c/2$, where
$\gamma_c/2\pi$ is the cavity linewidth), and using
$\Gamma=2\Gamma_A=8d^2_A\om^3_A/3\hbar c^3$, we obtain

\begin{equation}
\Gamma_c \approx
\left(1+\frac{3}{2}\frac{\Delta\Omega_{eff}}{4\pi}N\,\Lambda_{00}
\frac{\om_L^3}{\om_A^3}\right)\Gamma.
\label{gammac}
\end{equation} 

To verify whether this can be a significant enhancement of 
spontaneous emission, we introduce realistic experimental 
parameters. We consider a (quasi-)confocal cavity with mirrors of 
diameter $2b=1.0$~cm and reflectivities $r=0.97$ separated by 
$\ell=2.9$~cm. The excitation field, with circular  polarization, 
matches a TEM$_{00}$ mode. The cavity  linewidth is then 
$\gamma_c/2\pi\approx 200$~MHz and the line shape function is 
$\Lambda_{00}=66$, with a corresponding  finesse ${\cal F}=103$. 
We still need to determine the  number $N$ of entangled 
quasimolecules and consider their  distribution along the cavity 
mode. We assume that our  sample of cold $^{85}$Rb atoms 
($\lambda_A=2\pi c/\om_A =795$~nm and $\Gamma_A/2\pi = 6$~MHz) is 
trapped in a FORT (see Fig.~\ref{fig:2}), near the center of the cavity, in a 
cigar-shaped cloud with length $L\approx 0.6$~mm and radius 
$a\approx 2.6\times 10^{-2}$~mm. The number $N$ of quasimolecules 
in the state $|E;0\ra$, for a detuning $\delta=\om_L - \om_A = 
-2\pi\times100$~MHz, can then be estimated by counting all pairs 
such that $U(R_C)-U(R)\leq\hbar\Gamma$, since the excitation 
laser linewidth ($\sim 1$~MHz) is negligible. With 
$C_3=11.4\times10^{-11}$ erg\,\AA$^3$, this gives a spread 
$\Delta R\sim \hbar\Gamma/|U^\prime (R_C)|\sim 22.4$~\AA\ about 
$R_C\simeq 556$~\AA. For $N_A\sim 10^6$ atoms at a density $n_A 
\sim 10^{12}$~cm$^{-3}$ we have $N\simeq\frac{1}{2}N_A n_A 4\pi 
R_C^2 \Delta R\sim 45$ pairs. Even though this is a relatively 
crude approximation, attaining a number of this order should be 
feasible, since even larger numbers $N_A$ of atoms in a FORT were 
recently reported by Corwin {\em et al.}~\cite{corwin99}. 
Moreover, if the experiment were to be performed in a magnetic 
trap, $N_A$ could be larger by a few orders of magnitude. The  
distribution of quasimolecules in the cavity is simulated by 
sorting out 10 sets of 45 random positions ${\bf r}_i$ and dipole 
orientations ${\bf d}_i$. After averaging, we obtain 
$\Delta\Omega_{eff}/4\pi=7.4\times 10^{-4}$ and 
$\Gamma_c/\Gamma\approx 4.3$. 

Such an enhancement of spontaneous emission would lead to 
observable consequences, as in the process of radiative escape of 
atoms from the trap. This process does not occur naturally in a 
FORT, since the atoms are trapped in the ground state. It has to 
be induced by a separate probe laser, as in the photoassociation 
experiments by Cline {\em et al.}~\cite{miller93}. If this laser 
beam is sent via the cavity mode, an entangled state will be 
excited and the spontaneous emission rate will show collective 
enhancement. On the other hand, if it propagates perpendicular to 
the cavity axis (see Fig.~\ref{fig:2}), the interactions with 
individual quasimolecules will not be indistinguishable and the 
spontaneous emission rate will be at most cavity-enhanced, thus 
close to $\Gamma$ \cite{feld}. {\em We can therefore compare 
directly these two situations and measure trap losses ``with'' and 
``without'' cavity}. 

The trap-loss probability will be calculated using semi-classical 
models~\cite{weiner99,gp89,suominen98}, for which the probability 
of excitation is treated independently from emission. These 
models have been shown to give good results in the regime of low 
excitation laser intensities (we consider only a single 
excitation in the cavity) and for detunings $\delta \gtrsim 10 
\Gamma_A$~\cite{suominen98}. We describe the specific 
transition chosen here ($5S_{1/2}+5S_{1/2}\rightarrow 
5S_{1/2}+5P_{1/2}$ asymptotic states) as a two-level system. As shown 
by Peters {\it et al.}~\cite{peters94}, multi-level crossings are 
not important to this transition for our detuning range. The net result 
is an average over several similar hyperfine-splitted two-level 
systems, which would not lead to substantial 
modifications in the orders of magnitude we calculate. 
All these models deal only with the 
trap-loss probability of single quasimolecules but, as we show below, 
they can be adapted to consider an excited entangled state.  

In order to estimate trap losses, we consider a trap depth $V_0$ 
of the FORT of the order of 5~mK ($\sim 100$~MHz). The kinetic 
energy gained has then to be greater than 10~mK for both atoms of 
the quasimolecule to escape from the trap. This places an upper bound 
for the cavity linewidth $\gamma_c/2\pi\le 2V_0\sim 200$~MHz, that 
defines the resonant region $R_e<R<R_C$ where emission does not lead 
to trap loss. As the atoms of a 
colliding pair accelerate towards each other, the decay rate of
$|E;0\ra$ shifts from $\Gamma_c$ to the {\em off-resonance} value
$\Gamma$~\cite{details}, since $R$ for each component state $|i;0\ra$ of
$|E;0\ra$ decreases below $R_e$. After this first passage in the region 
$R<R_e$, the probability that two atoms, from any quasimolecule $i$, 
escape from the trap is $\sum_i|V_i/\hbar\tilde\Omega|^2\,
(1-e^{-2t_e\Gamma})e^{-t_c\Gamma_c}$, 
where $t_c$ ($t_e$) is the time interval spent between $R_C$ and $R_e$
($R_e$ and $R=0$). Notice that $\sum_i|V_i/\hbar\tilde\Omega|^2 = 
\la E;0|E;0 \ra =1$, so the difference from the expression obtained 
``without cavity'' is the existence of two radiative damping 
rates $\Gamma$ and $\Gamma_c$. The vibrational levels of 
$U(R)$ are accounted for by allowing multiple-passages
across $R_e$~\cite{peters94} before emission occurs (for our
detuning, these levels are not resolved and a wavepacket containing
several levels is excited~\cite{gp89,suominen98,peters94}). 
The overall loss-probability 
${\cal L}_c$ is then 

\begin{eqnarray}
{\cal L}_c & = & \sum_{i=1}^{N} \left|\frac{V_i}{\hbar\tilde\Omega}\right|^2\,
\sum_{n=1}^{\infty}(1-e^{-2\Gamma t_e})e^{-(2n-1)\Gamma_c t_c 
-2(n-1)\Gamma t_e} \nonumber \\ [0.3 cm] 
 & = & \frac{\sinh{(t_0-t_c)\Gamma}}
   {\sinh{[t_0+(\Gamma_c/\Gamma-1)t_c]\Gamma}} \equiv p\,{\cal L}_0,
\label{loss}
\end{eqnarray} 

\medskip
\noindent
where ${\cal L}_0\equiv\sinh{(t_0-t_c)\Gamma}/\sinh{t_0\Gamma}$ is the
cavity-free loss-probability and $t_0=t_c+t_e=3.0\times 10^{-8}$\,s ($\sim
\Gamma_A^{-1}$) is the total time interval from  $R_C$ to $R=0$, obtained from
conservation of energy $\mu\dot{R}^2/2 + U(R) = $ const~\cite{gp89}, 
neglecting the initial velocity $\dot{R}$ at $R_C$. Here $\mu$ is the reduced
mass of the colliding pair. Approximating
$\sinh(x)\approx e^x/2$ for $x\geq t_0\Gamma\approx 2.3$, we see that ${\cal
L}_c/{\cal L}_0=p\approx e^{-(\Gamma_c-\Gamma)t_c}$ is simply limited 
by the ratio between the survival probabilities, in the first passage through
the region $R_e<R<R_C$, with and without cavity. In order to measure 
${\cal L}_c/{\cal L}_0$, we can adjust the laser intensities
to have the same fraction of excited-state quasimolecules in both
situations (excitation via cavity mode or excitation beam
perpendicular to the cavity axis). The excitation probability factors
out, and we obtain with our parameters ($\Gamma t_c \approx 1.7$)  
${\cal L}_c/{\cal L}_0\approx 3\times 10^{-3}$. 

This large predicted trap loss suppression should encourage future attempts 
to observe it experimentally. Other experimental setups could 
have been considered, in which larger trap depths and numbers of 
atoms are obtained (such as magnetic traps), leading to even more 
encouraging orders of magnitude. Our results could also be 
extended to a situation with a larger fraction of excited 
molecules, closer to the superradiant
regime~\cite{dicke,bonifacio,superrad}. One must notice, however, that
this is not a method directly applicable to suppressing trap loss in a
MOT, for example. Quasimolecules {\it must} be excited via the cavity
mode, which would not happen in most existing optical traps. On the
other hand, trap loss may conceivably be prevented in novel traps
using optical resonators, as proposed in~\cite{vuleticchu}. 

In summary, we have shown, for the first time, that collisions 
between cold atoms can be manipulated by controlling the 
spontaneous emission time. This is achieved through  
multiparticle entanglement in a cavity-modified electromagnetic 
vacuum. Orders of magnitude of trap-loss probabilities show that this effect 
may be experimentally 
observed with present-day technology.

The authors acknowledge helpful discussions with K.L. Corwin, D. Kleppner, 
A. Lezama, R. Napolitano, H.M. Nussenzveig and M. Raizen, and financial 
support from FAPESP and CNPq.

\end{document}